\documentstyle[prl,aps,amssymb,graphicx]{revtex}
\newcommand{\beq}{\begin{equation}}
\newcommand{\eneq}{\end{equation}}

\input{epsf}

\begin{document}

\tolerance 10000

\twocolumn[\hsize\textwidth\columnwidth\hsize\csname %
@twocolumnfalse\endcsname

\draft

\title{Interaction between particles in a 1/$\lambda$-statistics anyon gas.}

\author {B. A. Bernevig$^{+}$, D. Giuliano$^\dagger$, D. I. Santiago$^*$
         }

\address{ $^{+}$ Department of Physics, Massachusetts Institute of Technology, 
        Cambridge, MA 02139\\
        $^\dagger$ Universit\`a di Napoli ``Federico II'' and 
        I.N. F.M., Unit\`a di Napoli, Monte S.Angelo - via Cintia, 
        I-80126 Napoli, Italy\\
        $^*$ Department of Physics, Stanford University,
        Stanford, California 94305}

%\twocolumn[
\date{\today}
\maketitle
\widetext

\begin{abstract}
%\vspace*{-1.0truecm}
\begin{center}

\parbox{14cm}{We analyze the dynamics between 1/$\lambda$-fractional statistics
particles (anyons) in an exact three-body solution of the Sutherland 
Hamiltonian. We show that anyons interact  by means of a short-range 
attraction. The interaction dictates important features of the spectral 
function for a charge-1 particle.}

\end{center}
\end{abstract}

\pacs{
\hspace{1.9cm}
}
]

\narrowtext

Particles constrained to move in a plane can exhibit
exchange statistics intermediate between Fermi statistics and Bose 
statistics\cite{frawil}. Intermediate-statistics particles are called anyons. 
Anyons occur in nature as the bulk excitations of  Fractional 
Quantum Hall (FQH) samples \cite{bobla}. In a 
FQH sample at filling fraction $\nu = 1/m$, the elementary charged excitation 
is an anyon whose total charge is $e/m$ and has $1/m$ statistics. A similar 
fractionalization of electric charge and statistics takes place within the 
edge excitations of such samples. 
Therefore, edges of a FQH-sample provide a physical realization of a
one-dimensional anyon gas \cite{cll}. 
Low-energy, long wavelength Quantum Hall edge excitations are described
by  a chiral Luttinger liquid model \cite{cll}. The Luttinger liquid 
formalism, however, cannot provide a  description of  processes with high 
momentum transfer between anyons (short-distance physics). Short-distance 
anyon physics is what we study in this paper, in an exact
solution of a simple one-dimensional model with 
$1/\lambda$-statistics excitations ($\lambda$ being a  positive integer 
number): the Sutherland model \cite{bills}. 
The Hamiltonian for $N$ particles lying at $x_1 , \ldots , x_N$
on a circle of length $L$ and constant density $\frac{N}{L}=1$ is:

\begin{displaymath}
H_S = \left( \frac{2 \pi}{ N} \right)^2 \biggl[ \sum_{i = 1}^N \left( 
z_i \frac{ \partial}{ \partial z_i } \right)^2 
\end{displaymath}
\beq
- \lambda (\lambda -1) \sum_{i \neq j = 1}^N \frac{ z_i z_j}{ ( z_i - z_j )^2} 
\biggr]
\label{one}
\eneq
\noindent
where $z_i = \exp[ \frac{ 2 \pi i x_i}{L} ]$.

\begin{figure}
\centering \includegraphics*[width=1.\linewidth]{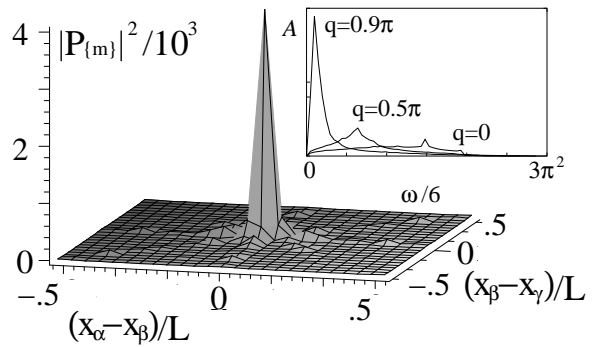}
\caption{Square of the three-anyon wavefunction versus the separation 
between anyons at $x_\alpha$ and $x_\beta$ ($x$-axis) and at $x_\beta$ and
$x_\gamma$ ($y$-axis). {\bf Inset}: spectral function vs. $\omega$ at $q=0.9 
 \pi$, $q=0.5  \pi$, $q=0$.}
\label{fig1}
\end{figure}

By generalizing the formalism developed in \cite{us1}, we construct and 
analyze the Schr\"odinger equation for any number of anyons. As a result, we 
show that anyons interact via a  short-range attraction.   When anyons are far 
apart, the probability of finding one at a certain position is essentially 
independent of  the positions of the others, showing that anyons are free when 
far enough away from each other. When
any number $M$ ($\le \lambda$) of anyons come together there is an increase of 
probability. The maximum enhancement of probability occurs when 
$\lambda$ of them come together in the $1/\lambda$ statistics anyon gas. The 
short 
distance probability enhancement is the progeny of the short distance 
attraction; anyon attraction largely enhances the probability of 
configurations with anyons at the same place.
Although the contribution to the total energy due to the interaction 
disappears in the thermodynamic limit, the enhancement does not. Instead, 
the peak value of the probability increases with $N$, and diverges in the
thermodynamic limit. The enhancement is the matrix element for a charge-1 
excitation to decay into $\lambda$-anyons. In the thermodynamic limit, 
the charge-1 excitation completely loses its integrity; as soon as 
a charge-1 particle/hole is created in the system, it immediately breaks up 
into $\lambda$ anyons.  In particular for the $\lambda=3$ case we display the 
relevant physics of anyon attraction in Fig.\ref{fig1} where we plot the square
modulus of the three-anyon wavefunction as a function of the separation between
two anyons at a time.

Let us start the mathematical derivation of anyon interaction.
The ground state of $H_S$ is

\beq
\Psi_{\rm GS} (z_1 , \ldots , z_N ) = \prod_{i < j =1}^N ( z_i - z_j )^\lambda
\prod_{ t = 1}^N z_t^{ - \lambda (N-1)/2}\; \; \; 
\label{two}
\eneq
with energy $E_0 = \left( \frac{2 \pi}{N}
\right)^2 \lambda^2 \frac{ N }{12} (N^2-1)$.   

Elementary excitations on top of such a ground state are the one-dimensional 
analogs of Laughlin's anyons for a $\nu=1/\lambda$ FQH-fluid. We create a 
$1/\lambda$ charge and statistics quasihole at $z_\alpha$ by removing 
$1/\lambda$ of an electron from the ground state. The excitation is a localized
charge defect given by the wavefunction:

\beq
\Psi_\alpha ( z_1 , \ldots , z_N ) =  \prod_{ j = 1}^N ( z_j - z_\alpha )
\Psi_{\rm GS} ( z_1 , \ldots , z_N ) 
\label{three}
\eneq
\noindent

$\Psi_\alpha$ is not an eigenstate of $H_S$. Energy eigenstates are 
provided by Fourier transforming $\Psi_\alpha$:

\beq
\Psi_m ( z_1 , \ldots , z_N ) = \oint \frac{ d z_\alpha}{ 2 \pi i ( z_\alpha
)^{1+m}}  \Psi_\alpha ( z_1 , \ldots , z_N ) 
\label{four}
\eneq
$(m = 0 , \ldots , N $ and $\oint$ denotes the integral over a closed
path surrounding the origin). 
The corresponding eigenvalues are given by $E_0 + E(q_m) = E_0 +  \left( 
\frac{2 \pi}{N} \right)^2 \left[ ( N - m ) \left( \lambda m + 1 \right) 
\right]$. The momentum of $\Psi_m$ is defined as $q_m \equiv \left( 
\frac{ 2 \pi}{N} \right) [ N / 2 - m ]$. In the thermodynamic limit $m/N$ 
is kept finite, and $E ( q_m )= \lambda  
\left[ \pi^2 - q_m^2 \right]$, where $-\pi \le q_m \le 
\pi$.

Multi-anyon states are constructed by a procedure analogous to the one
used for one-anyon states. The wavefunction for $M$ anyons localized at 
$z_{\alpha_1}, \ldots, z_{\alpha_M}$ is given by:

\beq
\Psi_{\alpha_1, \ldots, \alpha_M } ( z_1 , \ldots , z_N ) = 
\prod_{j = 1}^N \prod_{i = 1}^M( z_j - z_{\alpha_i })
\Psi_{\rm GS} ( z_1 , \ldots , z_N ) 
\label{five}
\eneq
\noindent
A convenient basis for the $M$-anyon eigenstates is given by Fourier
transforming Equation (\ref{five}):

\beq
\Psi_{m_1 , \ldots , m_M }  = \oint 
\Psi_{\alpha_1, \ldots, \alpha_M } \prod_{i = 1}^M 
\frac{ d z_{\alpha_i}}{ 2 \pi i z_{\alpha_i}^{1 + m_i} } 
\label{six}
\eneq
\noindent
with $0 \leq m_M \leq \ldots \leq m_1 \leq N$. 
These many-anyon plane waves are not energy eigenstates. However,
when acting on $\Psi_{m_1 , \ldots , m_M }$ with $H_S$, we
obtain \cite{bills}:

\[
[ H_S - E_0 ] \Psi_{ \{ m \}  } = \left( \frac{2 \pi}{N}
\right)^2 [   
 \sum_{i = 1}^M E(q_{m_i}) 
 - \sum_{i < j = 1}^M (m_i - m_j) ]  
\Psi_{ \{ m \}  }
\]

\beq
+ \sum_{ i< j= 1}^M
\sum_{k > 0 } (m_i - m_j + 2 k ) 
\Psi_{\ldots , m_i + k , \ldots , m_j      - k }
\label{seven}
\eneq
\noindent
In Eq.(\ref{seven}) $\{ m \} $ is shorthand for $ m_1 , \ldots ,
m_M$.

Eq.(\ref{seven}) is a lower-diagonal matrix equation. The energy eigenvalues
are the diagonal elements $E_{ \{ m \} } 
= E_0 +    \sum_{ i= 1}^M E(q_{m_i})
 - \left( \frac{2 \pi}{N} \right)^2 \sum_{i < j = 1}^M (m_i - m_j )    $. 
Many-anyon eigenstates of $H_S$,
$\Phi_{ \{ m \} }$, can be constructed from the $\Psi_{ \{ m \} }$'s by the 
so-called ``squeezing procedure'' \cite{bills}.
The $ (m_i - m_j ) $ terms make $E_{ \{ m \} }$
not be equal to the sum of the energies of the single isolated anyons. 
Therefore, these terms show an interaction for anyons with large relative 
momenta, i.e at short distances. The anyon interaction is attractive because of
the negative sign. 

In the thermodynamic limit, the energy eigenvalues $E_{ \{ m \} }$ 
becomes equal to

\beq
E_0 + \sum_{ i= 1}^M E(q_{m_i}) = E_0 + \lambda \sum_{ i= 1}^M (\pi^2 - q_{m_i}^2)
\label{nine} \; \; \; .
\eneq
\noindent
We see that the energy for an $M$-anyon eigenstate just reduces to the sum of 
the energies of the single isolated anyons. The interaction contribution 
apparently disappears  in the thermodynamic limit. 
However, as it happens with spinons \cite{us1}, the interaction plays a crucial
role in determining some of the relevant physics of the anyon gas, as we are 
going to discuss next.

The real-space wavefunction for $M$ anyons localized at $z_{\alpha_1}, \ldots,
z_{\alpha_M}$ can be expanded in terms of the energy eigenstates: 

\begin{displaymath}
\Psi_{\alpha_1, \ldots, \alpha_M} = \sum_{m_1=1}^N \sum_{m_2=0}^{m_1} \ldots
\sum_{m_M = 0}^{m_{M-1} }
\end{displaymath} 
\beq
\prod_{i=1}^M z_{\alpha_i}^{m_i}
{\cal P}_{ \{ m \} } \left(\frac{z_{\alpha_2}}{ z_{\alpha_1}} , 
\frac{z_{\alpha_3}}{ z_{\alpha_2}}, \ldots, \frac{z_{\alpha_M}}{ 
z_{\alpha_{M-1}}} \right) \Phi_{ \{ m\} }
\label{ten}
\eneq
\noindent
where the function ${\cal P}_{ \{ m \} } \left(\frac{z_{\alpha_2}}
{ z_{\alpha_1}},\frac{z_{\alpha_3}}{ z_{\alpha_2}}, \ldots, 
\frac{z_{\alpha_M}}{z_{\alpha_{M-1}}} \right)$ is a polynomial in each of its 
variables. By definition, $\prod_{i=1}^M z_{\alpha_i}^{m_i}{\cal P}_{ \{ m \} }
\left(\frac{z_{\alpha_2}}{ z_{\alpha_1}} , \frac{z_{\alpha_3}}{ z_{\alpha_2}}, 
\ldots, \frac{z_{\alpha_M}}{ z_{\alpha_{M-1}}} \right)$ is the
coordinate representation of the wavefuction for $M$ anyons in the 
state of energy $E_{ \{ m \} }$. 

The Schr\"odinger equation for  ${\cal P}_{ \{ m \} }$ is derived from
the  identity:

\[
\langle \Phi_{ \{m \} } | ( H_S - E_0 ) |  \Psi_{\alpha_1, \ldots, \alpha_M}
\rangle = ( E_{ \{ m \} } - E_0 ) 
\]

\beq
\times \langle \Phi_{ \{m \} } |  \Phi_{ \{m \} } 
\rangle
\prod_{i=1}^M z_{\alpha_i}^{m_i}
{\cal P}_{ \{ m \} } \left(\frac{z_{\alpha_2}}{ z_{\alpha_1}} , 
\frac{z_{\alpha_3}}{ z_{\alpha_2}}, \ldots, \frac{z_{\alpha_M}}{ 
z_{\alpha_{M-1}}} \right)
\label{eleven}
\eneq
\noindent
and from the fact that $H_S$ acts on $\Psi_{\alpha_1, \ldots, \alpha_M}$ as:

\[
 ( H_S - E_0 ) \Psi_{\alpha_1, \ldots, \alpha_M} = \left( \frac{ 2 \pi}{N} 
\right)^2
\biggl[ (\lambda N -2 ) \sum_{i=1}^M
z_{\alpha_i} \frac{ \partial}{ \partial z_{\alpha_i}}
\]

\[
-\lambda \sum_{i=1}^M \biggl( z_{\alpha_i} \frac{ \partial}{ 
\partial z_{\alpha_i}} \biggr)^2 + M^2 N
\]

\beq
-\frac{1}{2}\sum_{i \neq j=1}^M \biggl( \frac{z_{\alpha_i} + z_{\alpha_j}}
{z_{\alpha_i} - z_{\alpha_j}} \biggr) \left( z_{\alpha_i} \frac{ \partial}{ 
\partial z_{\alpha_i}} - z_{\alpha_j} \frac{ \partial}{ \partial z_{\alpha_j}} 
\right) \biggr] \Psi_{\alpha_1, \ldots, \alpha_M}
\label{twelve}
\eneq
\noindent

By putting together Eqs.(\ref{ten},\ref{eleven},\ref{twelve}), we obtain the 
equation of motion for the relative coordinate 
M-anyon wavefunction ${\cal P}_{ \{ m \} }\left(\frac{z_{\alpha_2}}
{ z_{\alpha_1}},\frac{z_{\alpha_3}}{ z_{\alpha_2}}, \ldots, 
\frac{z_{\alpha_M}}{z_{\alpha_{M-1}}} \right)$:

\[
\bigg\{ 2\lambda \sum_{i=1}^M m_i z_{\alpha_i} \frac{\partial}{\partial
z_{\alpha_i}}  + \lambda \sum_{i=1}^M  \biggl(
 z_{\alpha_i} \frac{ \partial }{ \partial z_{\alpha_i}} \biggr)^2   
\]

\[ 
+  \sum_{i < j =1}^M \biggl( \frac{z_{\alpha_i} + z_{\alpha_j}}
{z_{\alpha_i} - z_{\alpha_j}} \biggr) \left( z_{\alpha_i} \frac{ \partial}{ 
\partial z_{\alpha_i}} - z_{\alpha_j} \frac{ \partial}{ \partial z_{\alpha_j}} 
\right)
\]

\[
+ \sum_{i < j =1}^M (m_i - m_j)\biggl( \frac{z_{\alpha_i} + z_{\alpha_j}}
{z_{\alpha_i} - z_{\alpha_j}} \biggr)
\]
\beq
+ \sum_{i < j =1}^M (m_i - m_j) \biggr\} {\cal P}_{ \{ m \} } 
\left(\frac{z_{\alpha_2}}
{ z_{\alpha_1}},\frac{z_{\alpha_3}}{ z_{\alpha_2}}, \ldots, 
\frac{z_{\alpha_M}}{z_{\alpha_{M-1}}} \right)  = 0 \; \; \; .
\label{thirteen}
\eneq \noindent
Eq.(\ref{thirteen}) for three anyons and any $\lambda$ can be solved 
exactly \cite{us2}.
In order to illustrate the features of anyon interaction, in the rest of
this letter we shall analyze the full solution of Eq.(\ref{thirteen}) for the 
case of three anyons and $\lambda=3$ (the case of two spinons, i.e. two anyons 
with $\lambda=2$ has been solved in \cite{us1}.) 
$|{\cal P}_{ \{ m \} }(z,w)|^2 $, with  $z = z_{\alpha_2} / z_{\alpha_1}$ 
$w = z_{\alpha_3} / z_{\alpha_2}$, is the function we plot in Fig.\ref{fig1}, 
for $\lambda=3, N=20, m_1 - m_2 =10, m_2 - m_3=7$. 

We can extract physical consequences of
anyon attraction by  computing the spectral density for the charge-1
excitation. The state where one more charge-1 hole has been created 
at $\xi$, on top of the $N$-particle ground state, is described by the 
wavefunction
$\Psi_\xi (z_1 , \ldots , z_N ) = \prod_{j=1}^N ( z_j - \xi )^3 
\Psi_{\rm GS} ( z_1 , \ldots , z_N ) $. From the definition of ${\cal P}_{ \{
m \} }$ we find the following decomposition for $\Psi_\xi$ in terms of the
states $\Phi_{ \{ m \} }$:

\beq
\Psi_\xi = \sum_{m_\alpha = 0}^N \sum_{ m_\beta = 0}^{m_\alpha } 
\sum_{ m_\gamma = 0 }^{ m_\beta } \xi^{m_\alpha + m_\beta +  m_\gamma} 
{\cal P}_{ \{ m \}  }  ( 1 , 1 )
\Phi_{ \{ m \} } 
\label{ad1}
\eneq
\noindent
Since only positive-energy states propagate, the Green function for a charge-1
hole is defined as $G ( \xi , t ) = - i \theta ( t ) \langle \Psi_\xi (t )
| \Psi_{\xi=1} (t=0 ) \rangle$. The Green function correponds to a physical 
hole, which is obtained by pulling a real electron out of the system.
The spectral density of states, ${\cal A} 
( \omega , q )$ is the imaginary part of the Fourier transform of $G$ with 
respect to both $t$ and $\xi$, that is:

\[
{\cal A} ( \omega , q ) = \sum_{0 \leq m_\gamma \leq m_\beta \leq m_\alpha 
\leq N } \frac{ | {\cal P}_{ \{ m \} } ( 1 , 1 ) |^2 \langle \Phi_{ \{ m \}} |
\Phi_{ \{ m \} } \rangle }{ \langle \Psi_{\rm GS} | \Psi_{\rm GS} \rangle }
\]

\beq
\times
\delta_{m_\alpha + m_\beta + m_\gamma - n} \delta ( \omega - ( E_{ \{ m \} }
- E_0 ))
\label{specfu}
\eneq
\noindent
($q = 2 \pi n / L$, $0 \leq n \leq 3 N$  is the total momentum of the hole). 

In order to compute ${\cal A} ( \omega , q )$, we need the norm of the energy 
eigenstates and the enhancements ${\cal P}_{ \{ m \} } ( 1 , 1 )$. 
The norms are obtained by following the methods developed in \cite{us1} 
and can be expressed in terms of $\langle \Psi_{\rm GS} | 
\Psi_{\rm GS} \rangle$ = $ \Gamma [ 3 N + 1 ] / \Gamma [ 3 ]^N$ 
\cite{kwil}.
The solution is given by:

\[
\frac{ \langle \Phi_{ \{ m \}} | \Phi_{ \{ m \} } \rangle }{ \langle  
\Psi_{\rm GS} | \Psi_{\rm GS} \rangle } =
\prod_{i=1}^3 \frac{\Gamma [ N + 1 + (i-1)/ 3]}{\Gamma [ N +i/ 3]}
\times
\]

\[
 \prod_{i=1}^3 \frac{\Gamma [N - m_i + i/ 3] \Gamma[ m_i + (4-i)/
3 ]}{\Gamma [ N - m_i + 1 + (i-1)/ 3 ] \Gamma [ m_i + 1 + (3-i)/
3 ] } \times
\]

\[
 \prod_{i<j=1}^3 \frac{\Gamma [ m_i - m_j + (j-i)/ 3] 
\Gamma [m_i - m_j + (j-i)/3 + 1]}{\Gamma [ m_i - m_j + 1 + (j-i -1)/
3 ] \Gamma [m_i - m_j + (j-i + 1)/ 3 ]}
 \; \; . \]
${\cal P}_{ \{ m \} } ( 1 , 1 )$ can be computed from equation 
(\ref{thirteen}). The derivation will be described in a forthcoming
paper \cite{us2}. Here, let us just quote the main result:

\[
{\cal P}_{ \{ m \} } ( 1 , 1 )= \frac{ \Gamma^2 [ \frac{1}{ 
3 }]}{ \Gamma [ \frac{2}{ 3 }]}
\prod_{i<j=1}^3 \frac{\Gamma [ m_i - m_j + \frac{j-i +1}{3} ]}
{\Gamma [ m_i - m_j + \frac{j-i}{ 3 } ] }
\]
\noindent
By inserting  in Eq.(\ref{specfu})  the formulas for the  norms and for the 
enhancements, it is straightforward to work out 
${\cal A} ( \omega , q )$ in the thermodynamic limit. 
In this limit, we can trade the sums for integrals over the Brillouin zone
of the three anyons, so that Eq.(\ref{specfu}) takes the form:

\[
{\cal A} ( \omega , q ) = \frac{ \Gamma [ \frac{1}{3} ] }{
 \Gamma^2 [ \frac{2}{3} ] } 
\int_{-\pi}^\pi   \prod_{\rho=\alpha , \beta \gamma}
\left( \frac{ d q_\rho }{6} \right)   \; \delta [ \omega - 
E ( q_\alpha , q_\beta , q_\gamma )  ]
\]

\[
\times\left| \frac{ (q_\alpha - q_\beta )
( q_\beta - q_\gamma) ( q_\alpha - q_\gamma ) }{ ( \pi^2 - q_\alpha^2)
(\pi^2 - q_\beta^2 ) ( \pi^2 - q_\gamma^2 ) } \right|^\frac{2}{3} 
\delta  [ q - q_\alpha - q_\beta - q_\gamma ]
\]

In the inset of Fig.\ref{fig1}, we plot ${\cal A} (q , \omega )$ at different 
values of $q$. We recognize a fundamental feature of the anyon 
gas in that no definite Landau quasiparticle peak appears. This corresponds
to the lack of integrity of the charge-1 hole we discussed before.
More importantly, we can distinguish a well-defined sharp peak in the 
spectral function, whose height crucially
depends on $q$. Such a ``spectral enhancement'' is the enhancement in the
matrix element for a hole to decay in three anyons, ${\cal P}_{ \{ m \} }
(1,1)$. The dependence on $q$ reflects the dependence of ${\cal P}_{ \{ m \} }
(1,1)$
on the relative anyon momenta, the larger the relative momenta, the larger  
the enhancement. Therefore, the maximum enhancement should appear for values
of $q$ that make the differences among anyon momenta as large as possible, 
namely, when two anyons lie at opposite sites of the Brillouin zone ($q \approx 
\pm \pi$), 
while a third one has momentum $q \approx 0$. On the other hand, the
double constraint of momentum and energy conservation implies that the
number of three-anyon states in which the hole decays is proportional to the
size of the intersection between the plane $\sum_i q_i - q = 0$, the
sphere $\sum_i q_i^2 = 3 \pi^2 - \omega / 6$  and the 
Brillouin zone $ [ - \pi,\pi]^3$. Such a size is maximum when all three
anyons lie at the corners of the Brillouin zone, that is, $q_i\approx
\pm \pi$ $i = 1,2,3$. The two competing maximum
conditions discussed above imply that the maximum in the spectral function in
this system does not appear exactly at the emission threshold, as it
happens with spinons \cite{us1}, but at some intermediate value of
$\omega$, depending on $q$. 

Such a trend clearly appears from the plots in the inset of Fig.\ref{fig1} 
as a signature of the nonelementarity of the charge-1 excitation and of
the corresponding attraction among fractionally charged quasiparticles.

In the present work, the results of \cite{us1} have been generalized to 
excitations of arbitrary statistics $1/\lambda$. The main result is that 
fractionalized particles of any statistics always interact through a short 
distance attraction discovered in \cite{us1}. We conclude that short 
distance attraction is a generic feature of the dynamics of fractionalized
excitations. Such an attraction has profound consequences in the spectral 
function of the charge-1 hole. The peculiar distribution of spectral weight
should provide a possible experimental evidence of anyon attraction. 

In conclusion, 
it clearly appears that, because of its generality, our result 
should be a common feature of any anyon gas and, in particular, that it might 
provide a simple explanation of
some of the interesting and puzzling tunelling experimental findings on FQH
samples \cite{glattli,israelis}.

As a mathematical appendix we provide the explicit recursion relation
for the coefficients of ${\cal P}_{ \{ m \} }$. We assume a polynomial 
solution of the form 

\[
 {\cal P}_{ \{ m \} } ( z , w ) = \sum_{p , q \geq 0} b_{p,q}  z^p
w^q \; ; \; \;  {\cal P}_{ \{ m \} } ( 0 , 0 ) =1
\]
\noindent
where the summations over $p$ and $q$ stop at some finite indices. 
The boundary condition implies $b_{0,0} = 1$. 
Moreover, by inserting into Eq.(\ref{thirteen}) the polynomial expansion, it 
is straightforward to work out a recursion relation
for the coefficients $b_{p,q}$:

\[
\biggl[ ( \mu - p + q )( p+2) + ( \nu - q - 2 ) ( q + 2)  
+ \frac{ p + q + 4}{ 3} \biggr] b_{p+ 2 ,q + 2} = 
\]

\[
\biggl[ ( \mu -p - 1 + q) ( p + 2 ) + ( \nu - q - 1) ( q + 1 ) 
 -1 + \frac{ \nu +   q  - 2 p }{ 3} \biggr] 
b_{ p + 2, q + 1} 
\]

\[
+ \biggl[ ( \mu - p + q + 1) ( p + 1 ) + ( \nu - q - 2 ) ( q + 2 ) -1
+ \frac{ \mu +  p   - 2 q }{3} \biggr] 
 b_{ p + 1 , q +2 } 
\]

\[
- \biggl[ ( \mu - p + q - 1) ( p + 1 ) + ( \nu - q) q  
+ \frac{\mu + 2 \nu + 1 + p - 2 q }{ 3}\biggr]  
b_{ p + 1 , q }
\]

\[
 - \biggl[
 ( \mu - p + q + 1) p + ( \nu - q - 1) ( q + 1 )  + 
\frac{\nu + 2 \mu + 1 + q - 2 p }{ 3} \biggr] 
 b_{ p , q + 1 }
\]

\beq
 + \biggl[ \mu p   + \nu q + p q - p^2 
- q^2 + \frac{ 2 ( \mu + \nu ) - p - q}{
3} \biggr]  b_{ p , q  }
\label{nineteen}
\eneq
\noindent
where, if either $p$ or $q$ (or both of them) is $<0$, $b_{p,q}=0$, 
$b_{ \mu+k,j} = 0 $ for $k > j$, $b_{ \mu + \nu + i, q} = 0$ for $i > 0$.

This work has been conceived at a suggestion by R. B. 
Laughlin. His support and his advice to us in writing this paper have been
extremely precious. We kindly thank him. We acknowledge interesting 
discussions with A. Tagliacozzo, P. Sodano, G. Maiella and M. Berg\'ere.

\end{document}